\documentclass[useAMS,usenatbib]{mn2e}

\usepackage{graphicx}
\usepackage{amssymb}
\usepackage{longtable}
\usepackage{natbib}
\usepackage{rotating}
\usepackage{color, colortbl}
\definecolor{Gray}{gray}{0.9}
\bibpunct{(}{)}{;}{a}{}{,} 

\title[Studying the evolution of galaxies in compact groups]{Studying the evolution of galaxies in compact groups over the past 3 Gyr. I. The nuclear activity}
\author[T. Bitsakis et al.]{T. Bitsakis,$^{1}$\thanks{E-mail: tbitsakis@astro.unam.mx} D. Dultzin$^{1}$, L. Ciesla$^{2,3}$, Y. Krongold$^{1}$, V. Charmandaris$^{2,3,4}$, A. Zezas$^{2,5,6}$ \\
$^{1}$Instituto de Astronom\'ia, Universidad Nacional Aut\'onoma de M\'exico, A.P. 70-264, 04510 D.F., Mexico\\
$^{2}$Department of Physics, University of Crete, 71003, Heraklion, Greece\\
$^{3}$Institute for Astronomy, Astrophysics, Space Applications \& Remote Sensing, National Observatory of Athens, GR-15236, Penteli, Greece\\
$^{4}$Chercheur Associ\'e, Observatoire de Paris, F-75014, Paris, France\\
$^{5}$Harvard-Smithsonian Center for Astrophysics, Cambridge, MA 02138, USA\\
$^{6}$Foundation for Research and Technology - Hellas (FORTH), Heraklion 71003, Greece\\
}
\begin{document}

\date{date-here}

\pagerange{\pageref{firstpage}--\pageref{lastpage}} \pubyear{2015}

\maketitle

\label{firstpage}

\begin{abstract}
We present the first -- of a series -- study of the evolution of galaxies in compact groups over the past 3 Gyr. This paper focuses on the evolution of the nuclear activity and how it has been affected by the dense environment of the groups. Our analysis is based on the largest multiwavelength compact group sample to-date, containing complete ultraviolet-to-infrared (UV-to-IR) photometry for 1,770 isolated groups (7,417 galaxies). We classified the nuclear activity of the galaxies based on optical emission line and mid-infrared diagnostic methods, as well as using spectral energy distribution fitting. We observe a 15\% increase on the number of the AGN-hosting late-type galaxies found in dynamically old groups, over the past 3 Gyr, accompanied by the corresponding decrease of their circumnuclear star formation. Comparing our compact group results with those of local isolated field and interacting pair galaxies, we find no differences in the AGN at the same redshift range. Based on both optical and mid-IR colour classifications, we report the absence of Seyfert 1 nuclei and we attribute this to the low accretion rates, caused by the depletion of gas. We propose that the observed increase of LINER and Seyfert 2 nuclei (at low-z's), in the early-type galaxies of the dynamically young groups, is due to the morphological transformation of lenticular into elliptical galaxies. Finally, we show that at any given stellar mass, galaxies found in dynamically old groups are more likely to host an AGN. Our findings suggest that the depletion of gas, due to past star formation and tidal stripping, is the major mechanism driving the evolution of the nuclear activity in compact groups of galaxies.
\end{abstract}



\section{Introduction}
Active galactic nuclei (AGN) are considered some of the most dramatic and enigmatic phenomena in the evolution of galaxies. It is now believed that most galaxies in the Universe may host a supper-massive black hole (SMBH) in their nucleus \citep{Kormendy04}, thus stressing the importance of understanding the key mechanisms creating, fuelling and quenching them. Despite the diversity of the observed nuclear activities, there is currently a general consensus that they are all different expressions of the same phenomena. According to the AGN unification scheme \citep{Antonucci93, Urry95}, this diversity depends on our viewing angle. The SMBH is surrounded by an accretion disk, which in turn is also enclosed in an optically thick dusty torus. Depending on our position, we can either observe directly towards the central broad line region (BLR), where we may observe a Type 1 source (a.k.a Seyfert 1; Sy1), or the torus shall block our view towards the center, allowing us to observe only its reflection on the narrow line region (NLR) clouds, and therefore a Type 2 AGN (a.k.a Seyfert 2; Sy2). As a consequence, the expected spectrum of a Sy2 is dominated by narrow lines ($<$1000 km s$^{-1}$) of ionized metals (such as [OIII]$\lambda5007$ and [NII]$\lambda6584$), whereas Sy1s have both narrow and broad lines ($\gg$1000 km s$^{-1}$). 

More recently, however, a number of studies suggested that Sy1 and Sy2 hosting galaxies display important intrinsic, as well as environmental differences. \citet{Steffen03} showed that broad-line AGN are dominating the high X-ray luminosities, whereas narrow-line AGN the lower ones. (the so-called ``Steffen effect''). To interpret these results one should assume either a modification to the unified scheme, to include X-ray luminosities, or the complete segregation of Sy1 and Sy2 sources as two different categories of nuclear activity. Some optical and infrared studies revealed that Sy2 galaxies are more likely to display circumnuclear star formation than Sy1s \citep[i.e.][]{Dultzin94, Maiolino97, Gu01}, with more than 50\% of them exhibiting nuclear starbursts \citep{Cid01}. Galaxies hosting a Sy2 nucleus, though, were found more frequently in interaction with close neighbours ($\le$100 kpc) than Sy1s or non-active galaxies \citep{Dultzin99, Krongold02, Koulouridis06, Villarroel14}. On the other hand, \citet{Wu09} showed that there is no statistical difference in the strength of the polycyclic aromatic hydrocarbons (PAHs), which probe star formation, between Sy1's and Sy2's. In addition, \citet{Elitzur12} proposed that Sy1 and Sy2 AGN are simply the extremes of the AGN distribution, where in the first case we observe directly towards the BLR whereas in the other dust blocks completely our view towards it. \citet{Tran01} showed that 50\% of the Sy2's in their sample displayed BLRs observed in reflected polarised light (a.k.a hidden-BLRs; HBLRs), confirming their obscured Sy1 nature. However, the remaining non-HBLR galaxies appear to have lower AGN luminosities, indicative of their lower accretion rates \citep{Wu11}. All the above results, raised concerns about the validity of the Unification scheme. Alternative models \citep[i.e.][and references therein]{Nicastro00,Nicastro03} suggested that the accretion rate, and subsequently the AGN luminosity, play a key role in the presence of the BLR. \citet{Elitzur06} showed that for low bolometric luminosities ($<$10$^{42}$erg sec$^{-1}$), the torus and possibly the BLR disappear, due to the inability of the SMBH to sustain the required cloud outflow rate. 

The existence of a connection between galaxy interactions and nuclear activity, both in terms of star formation and/or AGN activity, has been also confirmed by \citet{Storchi08}. Since most galaxies in the Universe are found in interacting environments -- such as pairs, groups and clusters \citep[e.g.][]{Small99} -- it is very important to understand their connection with the various types of nuclear activity. Although clusters host the largest number of galaxies, of all the interacting environments, it has been recently shown that a large number of their members have been pre-processed in groups \citep[i.e][]{Cortese06, Eckert14}. \citet{Hickson92} showed that compact groups (CGs) of galaxies appear as the ideal systems to study the effects of galaxy interactions, due to their high galaxy densities and low velocity dispersions ($\sim$250 km s$^{-1}$). So far, the most complete and best studied CG sample was {\bf the one} compiled by \citet{Hickson82}. The initial sample consisted of 100 CGs -- the so-called Hickson compact groups (HCGs), containing 451 galaxies -- however later \citet{Hickson92}, using spectroscopic information, reduced it to 92 groups having at least 3 accordant members. Although, the dynamical and star formation properties of these groups have been extensively examined during the last three decades \citep[i.e][]{Mendes94,Johnson07,Bitsakis10, Bitsakis11}, there is a limited number of studies focused on their nuclear activity. \citet{Shimada00} studied a sample of 69 galaxies belonging to 31  HCGs and found that nearly 40\% host an AGN. Later, \citet{Martinez08} and \citet{Martinez10} relied on the optical spectroscopy of 270 galaxies in 64 CGs, and confirmed the fraction of AGN previously found (42\%). They also showed that the majority of AGN in CGs have low luminosities (LLAGN), and they attributed this to their high gas deficiencies. This result is consistent with those of \citet{Verdes01}, \citet{MartinezBadenes12} and \citet{Bitsakis14}, where it was shown that HCG galaxies display high deficiencies in their atomic and molecular gas, as well as their dust content due to tidal stripping. More recently, \citet{Sohn13} studied a sample of 58 local CGs selected from the SDSS data release 7. They found AGN activity in 17-42\% of the galaxies, depending on the classification method used. They also showed that no powerful dust-obscured AGN were detected, according to the most recent WISE mid-IR classification method of \citet{Mateos12}, and they suggested that nuclear activity is not strong due to gas depletion. Finally, \citet{Tzanavaris14} examined $Chandra$ X-ray maps of 9 HCGs and showed that the majority of the galaxies displaying low specific star formation rates (thus being earlier types), are also more likely to host a weak AGN, with L$_{X, 0.5-8.0 keV} < 10^{41}$ erg sec$^{-1}$.

Although these studies examined the incidence of the AGN activity in local CG galaxies (z$<$0.05) and its connection with the dense environment of the groups, they were not able to show how it evolved throughout cosmic time. Nevertheless, the well known connection between star formation and the AGN activities \citep[e.g.][]{Magorrian98} and the fact that the star formation history of the Universe changed significantly over the past few Gyr \citep[e.g.][]{Heavens04}, imply that AGN activity should have also experienced several different phases, during this period. Using the advent of wide-area extragalactic surveys, such as the Galaxy Evolution Explorer surveys ($GALEX$), the Sloan Digital Sky Survey ($SDSS$), the 2 micron all-sky survey ($2MASS$), and the Wide-field Infrared Survey {\bf ($WISE$)} Explorer All Sky Release, as well as the published CG catalogues of \citet{McConnachie09}, we obtained a working sample of 1,770 CGs, containing 7,417 galaxies, in the redshift range of 0.01$<$z$<$0.23 (thus, looking back time of approximately 3 Gyr). Using this sample, we will examine the evolution of the AGN activity and its connection with the dynamic environment of the groups. In \S2, we describe the selection of our sample, and also the comparison samples we have used to put the properties of our galaxies into context. In \S3, the nuclear and morphological classifications are presented, as well as the spectral energy distribution modeling we performed to estimate some important physical properties of the galaxies, necessary for our study. In \S4, we present results, and in \S5 we discuss their implications. Finally, in \S6, we sum up our main conclusions.

To calculate the distances in this paper we adopt a flat $\Lambda$CDM cosmological model, with parameters: H$_{0}$=70 km s$^{-1}$ Mpc$^{-1}$, $\Omega_{m}$=0.30 and $\Omega_{\Lambda}$=0.70.

\section{The samples}

\subsection{The SDSS compact group sample selection}
The sample of compact groups of galaxies, presented in this paper is part of the larger sample of \citet{McConnachie09}. The latter was constructed by applying the slightly-improved Hickson's criteria, to the whole Sloan Digital Sky Survey Data Release 6 ($SDSS$-DR6; \citealt{Adelman08}). \citet{Hickson82} defined as compact all galaxy groups in which, $(i)$ the number of galaxies within 3 magnitudes are at least four, N($\Delta$m$_{r}$=3)$\ge$4, $(ii)$ the angular diameter with no additional galaxies is at least 3 times the diameter of the group, $\theta_{N}\ge 3 \theta_{G}$, and $(iii)$ the total magnitude of the galaxies, averaged over $\theta_{G}$, is less than 26mag arcsec$^{-2}$. However, \citet{McConnachie08} showed that by increasing the required minimum surface brightness, from 26-to-22 mags arcsec$^{-2}$, the contamination of groups containing an additional non-member galaxy decreases from 71\% to 26\%. Moreover, they imposed a faint-end limit at Petrosian r-band magnitude of 18 mags, to ensure that the observational and the mock catalogues (created to test the selection criteria) are robustly compared, and a bright-end limit of 14.5 mags, to ensure that automatic de-blending of galaxies is reliable. Their final catalogue comprised 2,297 compact groups, containing 9,713 galaxies (see Catalogue A of \citealt{McConnachie09}). For the purposes of this work we have constructed a sub-sample of the Catalogue A of \citet{McConnachie09}, using as a criterion the availability of ultraviolet (UV), optical, near and mid-infrared (IR) photometry, which are essential to estimate the physical properties of the galaxies and perform a detailed analysis of the sample. 

Initially, since Catalogue A originates from the $SDSS$, optical photometry was available for all the galaxies. The $SDSS$-DR7 \citep{Abazajian09} covered over 35\% of the sky in the $u, g, r, i, z$ bands, centered at 3557, 4825, 6261, 7672 and 9097$\AA{}$, with a resolution of 1.4 arcsec and sensitivities ranging from 22.0-to-20.5 mags for 5-$\sigma$ detection, respectively. The photometry was automatically estimated using a combination of two models, a pure de Vaucouleurs profile and a pure exponential profile, which is considered the optimal for extended sources (called the CModel Magnitudes; \citealt{Strauss02}). Simultaneously, spectroscopy was performed at all sources brighter than 17.77 mags in r-band,  with angular separation more than 55 arcsec to avoid collision of the $SDSS$ fibers. The size of the $SDSS$ aperture was 3\arcsec (see more in \S3.4). The fluxes and equivalent widths of the optical lines have been obtained from the work of \citet{Brinchmann04}. These authors have published the processed line fluxes of more than 8$\cdot$10$^{5}$ galaxies from the $SDSS$-DR7 (see MPA-JHU DR7\footnote[1]{http://www.mpa-garching.mpg.de/SDSS/DR7}). They fitted single Gaussian profiles after removing the stellar continuum, correcting for the foreground galactic reddening and re-normalising them to match the photometric fiber magnitudes in the r-band. In addition, they re-estimated the uncertainties to include both errors in spectrophotometry and continuum subtraction uncertainties. Unfortunately, they have chosen 500 km s$^{-1}$ as an upper limit of the width of the Gaussian profile, to ensure the correct estimation of the faintest line fluxes where widths are often overestimated, thus preventing us from classifying real broad-line galaxies (with widths $>$1000 km s$^{-1}$). The fraction of the emission line galaxies in our sample, having signal-to-noise ratios S/N$>$3, was 75\%.

For the near-IR observations we have used the published magnitudes from the $2MASS$ extended source catalogue \citep{Skrutskie06}. The galaxies were imaged in J, H, and K bands, with 3-$\sigma$ sensitivity limits of 14.7, 13.9, and 13.1 mags, respectively. The photometry was performed using sets of circular and ellipsoidal apertures applied on the 3-$\sigma$ J-band isophote of each individual galaxy, so that both the integrated flux, as well as the estimated uncertainty were estimated. 

Using the latest $WISE$ \citep{Wright10} all sky release, we obtained the photometry or the upper limits for the galaxies in our sample. $WISE$ was a 40cm telescope, which performed all sky survey, covering the 3.4, 4.6, 12.0, and 22.0$\mu$m bands (named w1-to-w4), with 5-$\sigma$ sensitivity limits 19.70-to-14.40 mags, respectively. The $WISE$ photometry (WPHOT) was automatically performed using profile fitting. However, if the PSFs of two or more galaxies were closer than the resolution of the telescope (6-to-12 arcsec), WPHOT reports the  total photometry with a blending parameter, which is the number of PSFs included in the aperture.

The UV data were obtained from $GALEX$ \citep{Morrissey05} All Sky Survey (AIS; with 100sec on-source exposures), Medium Imaging Survey (MIS; with 1500sec on-source exposures), Nearby Galaxy Survey (NGS; with typical exposures of 4000sec), Deep Imaging Survey (DIS; with typical exposures of 4000sec), as well as from Guest Investigator's data 1-to-4 (GI1-4; with typical exposures of 1500sec); publicly available in the $GALEX$ archive. $GALEX$ is a 50cm diameter UV telescope that imaged the sky simultaneously in both FUV and NUV channels,  centered at 1540\AA{} and 2300\AA{}, respectively. The field-of-view (FOV) is approximately circular  with a diameter of $1.2^{o}$ and a resolution of about 5.5$''$ (FWHM) in the NUV. The data sets used in this paper are based-on the $GALEX$ sixth data release (GR6). The photometry was initially acquired from the GALEX archive, where it was automatically performed using SExtractor \citep[see][]{Bertin96}, a code that estimates the fluxes of all the sources in a given image using ellipsoid apertures. 

\begin{figure}
\begin{center}
\includegraphics[scale=0.52]{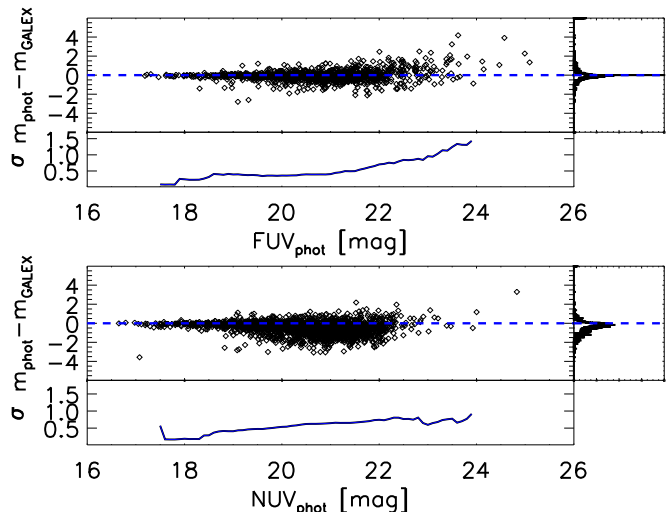}
\caption{Differences between the $GALEX$ photometry of the galaxies in our sample, derived manually using aperture photometry (indicated as $phot$) and the archival ones (indicated as $GALEX$) in the FUV (top panel) and NUV (bottom panel) bands. In the bottom boxes are presented the standard deviations ($\sigma$) of the values created as running averages, in bins of 0.1 mag. (A coloured version of this figure is available in the online journal)}
\label{fig_galex}
\end{center}
\end{figure}

To check the accuracy of the archival photometry obtained in the various bands (from the UV-to-IR), we randomly selected a sample of galaxies and performed aperture photometry, using the galaxy 3-$\sigma$ isophotal contours. Our results suggested that the published $SDSS$, $2MASS$ and $WISE$ archival fluxes were consistent with our photometry, within the uncertainties. However, for more than 10\% of the $GALEX$/NUV fluxes and for $\sim$34\% of the $GALEX$/FUV fluxes we found differences of 1-2 magnitudes (see Fig.~\ref{fig_galex}). To overcome this limitation in the UV fluxes, we manually performed aperture photometry by calculating the isophotal contours around each source in order to account for variations in the shape of the emitting region. Then we defined a limiting isophote 3-$\sigma$ above the local overall background, for each galaxy, and we measured the flux within this region, after subtracting the corresponding sky. To convert from counts to UV fluxes we used the conversion coefficients, given in the header of each file. The upper limits depend on the survey ranging from 20.5 and 21.5 mags, to 24.0 and 24.5 mags, for FUV and NUV bands respectively.

Our final sample comprises 1,770 groups, containing 7,417 galaxies with available UV-to-mid-IR flux densities. Due to $SDSS$ fiber collision constraints, optical spectroscopy was available only for 4,208 of these galaxies (for the purposes of this paper, galaxies with no spectral information were assigned with the z of the most massive galaxy in their group). This sample is nearly 55 times larger than the previous HCG multiwavelength samples presented in \citet{Bitsakis11, Bitsakis14}. It also covers a much  larger volume in the Universe, with groups being detected up to a redshift of 0.23, equal to a distance of $\sim$1.15 Gpc (the corresponding for HCGs is z$\sim$0.022 or 95 Mpc). 

\subsection{The comparison samples}
As comparison samples we have used truly isolated galaxies in the field, as well as galaxies in isolated interacting pairs. The first sample contains 513 local (with z below 0.05) isolated field galaxies presented in \citet{Hernandez13}. It mostly consists of late-type galaxies (86\%), with an AGN fraction of 40\%, at least for the 87\% of them for which optical emission lines could be measured. The second, is a sample of 385 isolated interacting pair galaxies, 71\% of which are LTGs \citep[originally selected from the catalog of][]{Karachentsev72}, was taken from \citet{Hernandez14}. Approximately 82\% of these galaxies have emission lines, and $\sim$49\% of them are classified as AGN-hosting galaxies. The authors separated the pairs into three subsamples depending on the morphologies of the member galaxies. Pairs consisting of late-type galaxies (S+S), of early-type galaxies (E+E), and mixed pairs (S+E). Both samples were selected from the $SDSS$-DR7 survey, having the same spectroscopic limitations than our compact group sample. 

\section{Data analysis and classifications}

\subsection{Morphological classification}
\begin{figure}
\begin{center}
\includegraphics[scale=0.52]{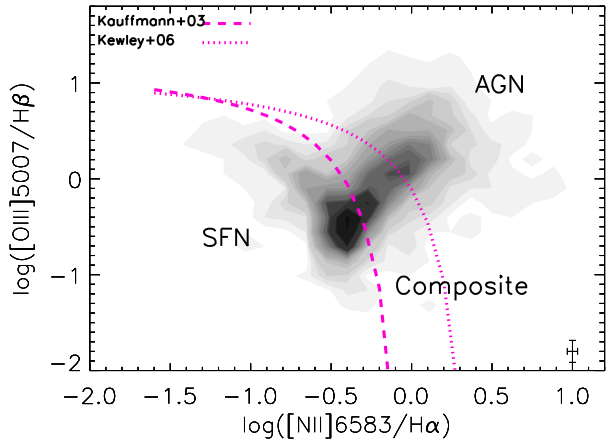}
\caption{The BPT diagnostic diagram of 4,208 galaxies in our sample with available optical spectroscopy. The dashed and dotted lines indicate the \citet{Kauffmann03} and \citet{Kewley06} AGN selection criteria, respectively. Galaxies located under the dashed line are classified as star forming nuclei (SFN), galaxies over the dotted line as pure AGN, and galaxies between the two as composite objects. Contours are at 5\% intervals of the maximum of the total distribution. Mean uncertainties are indicated in the bottom right corner and correspond to the 6\% and 16\% of the x and y-axis values, respectively. (A coloured version of this figure is available in the online journal)}
\label{fig2}
\end{center}
\end{figure}

A very important step to put the properties of our galaxies into context, was the classification of their morphologies, as well as of the dynamical state of the groups (as proposed by \citealt{Bitsakis10}). To classify our galaxies into late-type (spirals and irregulars; hereafter LTGs) and early-type (lenticulars and ellipticals; hereafter ETGs), we have relied on the criteria described in \citet{Simard09, Simard11}. These authors have fitted the $SDSS$ r-band 2D, point-spread-function-convolved, bulge$+$disk decompositions of a sample of 1.12 million galaxies from the SDSS-DR7. Using four different decomposition procedures they ensured more robust structural parameters of the galaxies, even in crowded environments. According to their findings, galaxies with bulge-to-disk ratios B/T$\ge$0.35 and image smoothness at half-light radius S2$\ge$0.075 are classified as ETGs. Applying this classification in our sample, 3,045 galaxies are classified as LTGs (41\%), 4,367 as ETGs (59\%). There is no available classification for 5. The results are similar to what we have observed in HCGs (54\% LTGs and 46\% ETGs, respectively). 

As mentioned above, in \citet{Bitsakis10} we have proposed a dynamical-evolution classification for the groups, depending on their ETG fraction. There we showed that this kind of classification is physical and also consistent with previous classifications based on the distribution of the HI gas content of the groups \citep[see][]{Verdes01, Borthakur10}. According to this, groups with more than 25\% of early-type members are classified as ``dynamically old'' (DO), whereas groups with less or no ETGs as ``dynamically young'' (DY). Applying this classification to our current sample, we find 373 dynamically young (21\%), and 1,397 dynamically old groups (79\%). 

\begin{figure}
\begin{center}
\includegraphics[scale=0.85]{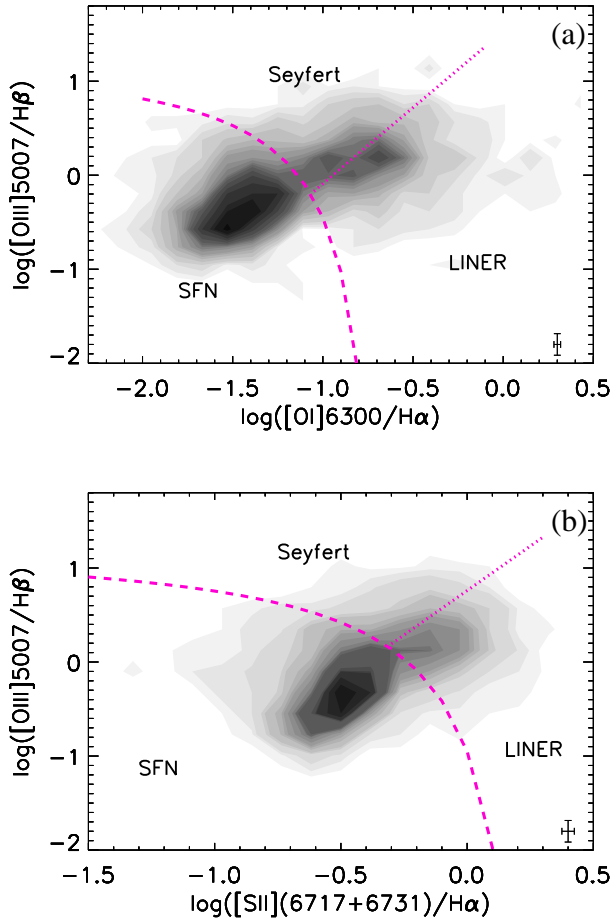}
\caption{The \citet{Kewley06} diagnostic BPT diagrams, that use [O{\sevensize I}]$\lambda6300$ (top panel) and [S{\sevensize II}]$(\lambda6717+6731)$ lines (bottom panel), to further separate AGN galaxies into Sy2 and LINERs (found in the upper left and lower right portion of both diagrams, respectively). Contours are at 5\% intervals of the maximum of the total distribution. Mean uncertainties are indicated in the bottom right corner of each panel and correspond to the 18\% and 16\%, as well as the 6\% and 16\% of the x and y-axis values of the top and bottom panels, respectively. (A coloured version of this figure is available in the online journal)}
\label{fig3}
\end{center}
\end{figure}

\subsection{AGN classification methods}
To characterise the nuclear activity of the galaxies in our sample, we have relied on the spectral classification diagnostic diagrams, initially proposed by \citet{Baldwin81} and later expanded by \citet{Veilleux87}, also known as BPT-diagrams. These, compare the flux ratios of several narrow emission lines, such as [O{\sevensize III}]$\lambda5007$/H$\beta$, [N{\sevensize II}]$\lambda6584$/H$\alpha$, [S{\sevensize II}]$(\lambda6717+6731)$/H$\alpha$ and [O{\sevensize I}]$\lambda6300$/H$\alpha$, and distinguish the nuclear activity of galaxies between star-forming nuclei (SFN), active galactic nuclei (AGN), and composite objects (objects showing both signatures of both SFN and AGN). In Fig.~\ref{fig2} we present the [O{\sevensize III}]$\lambda5007$/H$\beta$ versus the [N{\sevensize II}]$\lambda6584$/H$\alpha$ line ratios of the 4,208 galaxies in our sample with available optical spectroscopy \citep[see][]{Brinchmann04}. Following the selection criteria introduced by \citet{Kauffmann03} and \citet{Kewley06}, 1,436 of them are classified as SFN (34\%), 1,860 as AGN (44\%), and 912 as composite objects (22\%; detected to host both star formation as well as AGN activity within our aperture).

However, the above methods are not able to distinguish between broad-line Sy1, which appear to have very wide spectral lines with widths $>$1000 km s$^{-1}$, form narrow-line Sy2 sources. Nevertheless, as we explained in \S2.1, \citet{Brinchmann04} set an upper limit (of 500 km s$^{-1}$) to the width of the Gaussian profiles they have used to fit the spectral lines, to better constrain the weak narrow lines. Therefore, using these data, we have classified as ``broad-line AGN'', all AGN-hosting galaxies detected with signal-to-noise ratios $>$10$\sigma$, and lines wider than the upper limit, $\sigma>$500 km s$^{-1}$. This had as a result to include both Sy1s, as well as other narrower sources (having spectral line widths between 500 and 1000km s$^{-1}$). Yet, even applying this approximate classification, the fraction of ``broad-line'' AGN in our sample is negligible ($<$1\%).

\begin{figure}
\begin{center}
\includegraphics[scale=0.52]{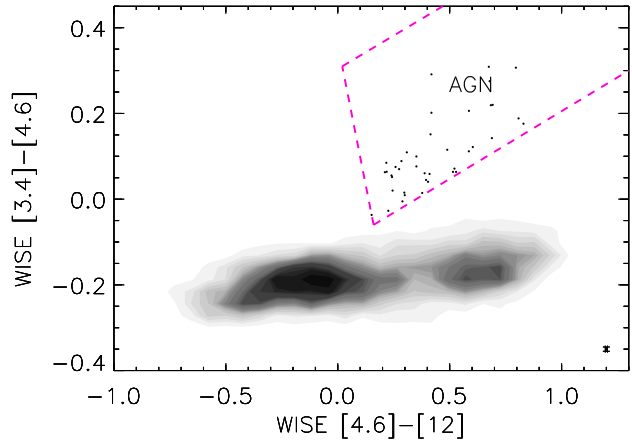}
\caption{The \citet{Mateos12} WISE [4.6]-[12] versus [3.4]-[4.6] colour diagram. Galaxies containing dust obscured AGN are located in the area within the dashed lines. Contours are at 5\% intervals of the maximum of the total distribution. Since, $<$1\% of our sources are found in the AGN locus, we identify them with black points. Mean uncertainties are indicated in the bottom right corner and correspond to the 4\% of the x and y-axis values. (A coloured version of this figure is available in the online journal)}
\label{fig5}
\end{center}
\end{figure}

More recent classifications were also able to distinguish between Sy2s and low-ionisation nuclear emission-line regions {\bf (LINERs; their classification as AGN is still open to discussion; e.g. see \citealt{Heckman80, Ho93})}. In the two panels of Fig.~\ref{fig3}, we present the diagnostic methods proposed by \citet{Kewley06}. According to these, we plot the [O{\sevensize III}]$\lambda5007$/H$\beta$ versus the [O{\sevensize I}]$\lambda6300$/H$\alpha$ (top panel) as well the [S{\sevensize II}]$(\lambda6717+6731)$/H$\alpha$ (bottom panel) line ratios, of all the galaxies detected in the classic BPT as AGN-hosting. Based on these classifications, 319/531 of  these galaxies are classified to host a Sy2 nucleus (8-13\%), whereas 685/851 are LINERs (16-20\%), respectively.

Finally, by applying the colour diagnostic methods of \citet{Mateos12} and \citet{Stern12} we were also able to identify dust obscured AGN activity. These methods compare the mid-IR colours of galaxies, to identify a power-law-based selection of luminous AGN candidates. The first method compares the WISE [4.6$\mu$m]-[12$\mu$m] versus the [3.4$\mu$m]-[4.6$\mu$m] colours (see Fig.~\ref{fig5}) and defines an area where power-law dominant AGN should be located. The second, describes different colour selection, where mid-IR luminous AGN should be located, at [3.4$\mu$m]-[4.6$\mu$m]$\ge$0.8 (Vega mags). Both methods predict the fraction of dust obscured AGN to be $<$1\%.

\subsection{Spectral energy distribution modeling}
\begin{figure}
\begin{center}
\includegraphics[scale=0.59]{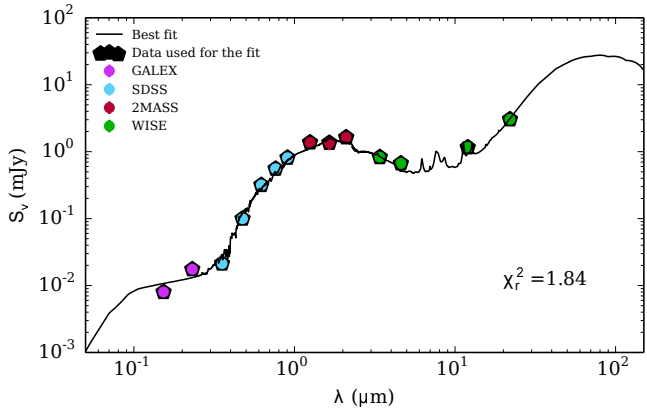}
\caption{Example of fits obtained with the {\sevensize CIGALE} code. The color code of the data-point refers to the instrument used, in purple: UV data from $GALEX$, in blue: optical data from $SDSS$, in red: NIR data from $2MASS$, and in green: MIR data from $WISE$. The black solid line shows the best fit model for each source. The value of the reduced $\chi^2$ is indicated for each fit. (A coloured version of this figure is available in the online journal)}
\label{figSED}
\end{center}
\end{figure}

We performed a UV to mid-IR broad band spectral energy distribution (SED; see Fig.~\ref{figSED}) fitting of all of the sources of the catalogue using the recently updated version of the {\sevensize CIGALE} model, which includes a detailed treatment of the AGN influence to the observed SED \citep[][Boquien et al. in prep, Burgarella et al. in prep]{Noll09, Ciesla14}. Based on an energy balance between the energy absorbed in UV-optical and re-emitted in IR, {\sevensize CIGALE} builds SED models that are then compared to the data. For each model, {\sevensize CIGALE} computes the $\chi^2$ value in order to build the probability distribution function (PDF) of each parameter. The final value of a parameter is thus the mean value of the PDF and the uncertainty associated is the standard deviation of the distribution.\\

We used a delayed star formation history (SFH) for our galaxies as it has been shown that it provides a good estimate of the stellar mass and SFR, as well as realistic ages \citep{Ciesla14}. This SFH was convolved with the stellar population models of \cite{Maraston05}. The SED was then attenuated using the \cite{Calzetti00} law, and the energy absorbed is re-emitted in the IR through the \cite{Dale14} templates. {\sevensize CIGALE} models the presence of an AGN through the library of \cite{Fritz06}, which takes into account the emission of the central object, the dust torus emission, and the scattered photons. However, when using broad band photometry, the AGN models are highly degenerated. Therefore, following \cite{Ciesla14}, we used three models: a Sy1, a Sy2, and an intermediate type AGN template, displaying a power law emission in mid-IR without any strong UV contribution. Two of the most important output parameters, for our current study, provided by {\sevensize CIGALE} are the stellar masses as well as the contribution of the AGN to the total IR luminosity, $f_{AGN}$, for each galaxy. We are confident in the presence of an AGN when $f_{AGN}$ is higher than 10\%. Between 5 and 10\% \citet{Ciesla14} showed that this confidence is reduced (yet the $\chi^{2}$ value is still 50\% lower than by not using the power-law template), and for fractions below 5\% we cannot conclude by the fit whether an AGN is present.

\subsection{Bias control and mass selection}

\begin{figure}
\begin{center}
\includegraphics[scale=0.54]{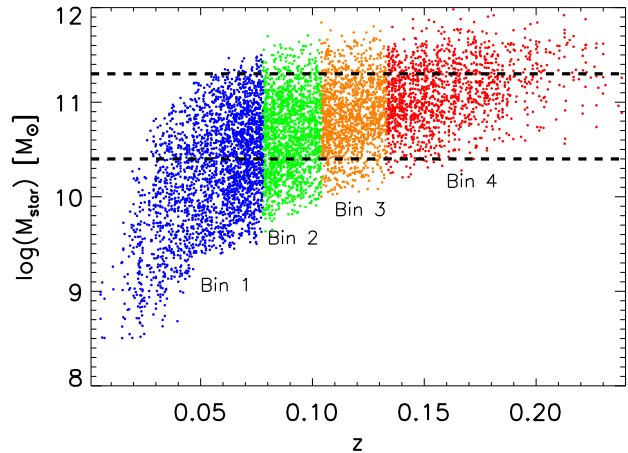}
\caption{Selection biases in our sample. Malmquist bias prevents us from observing less massive galaxies at higher redshifts. Moreover, due to an upper limit cut-off in the luminosities, during the selection process, brighter galaxies at lower-z's were also excluded. Galaxies selected in the stellar mass range of 10.4$\le$log(M$_{star}$)$\le$11.3 M$_{\odot}$, are located within the dashed lines. For the purposes of comparison we separate our sample into four different redshift regions, containing a similar number of galaxies ($\sim$1,100). These are: Bin1 (blue; 0.010$\le$z$<$0.078), Bin2 (green; 0.078$\le$z$<$0.104), Bin3 (orange; 0.104$\le$z$<$0.133) and Bin4 (red; 0.133$\le$z$<$0.230). (A coloured version of this figure is available in the online journal) }
\label{fig1}
\end{center}
\end{figure}

Due to its large size, with galaxies found in a wide range of redshifts and stellar masses, our sample can be subject to selection biases. In an effort to detect and control them, we plot in Fig.~\ref{fig1} the stellar masses of the galaxies in our sample as a function of their redshift. The Malmquist effect appears at higher redshifts (see lower-portion of this figure). {\bf Since}, we are unable to detect the faintest sources and, as a consequence we are biased towards the more massive ones. Furthermore, on the upper left portion of the same figure we can notice one more bias. As we mentioned in \S2.1, an upper limit cut-off was applied to the SDSS luminosities of our galaxies, to ensure the success of the automatic de-blending \citep[see][]{McConnachie09}. This had as a result to exclude the lower-z bright-massive galaxies from our sample. To overcome these two biases, while comparing galaxies at different redshifts, we chose objects within the mass range of 10.4$\le$log(M$_{star}$)$\le$11.3 M$_{\odot}$. This mass selection can also ensure that we are not including dwarf galaxies in our study, which could affect our statistics. Indeed,  the lower mass limit of log(M$_{stellar}$) = 10.4 M$_{\odot}$ refers to $SDSS$ z-band magnitudes brighter than -21.5, which correspond to bright galaxies, according to the $SDSS$ z-band luminosity functions presented in \citet{Blanton01}.

Moreover, for the purposes of comparison we have also separated our sample into four different redshift bins, chosen to contain a similar number of galaxies after the mass selection was applied (shown in Fig.~\ref{fig1} in different colours). In Table~\ref{tab_bin}, we present the number and the median stellar mass of the galaxies in each redshift bin, before and after the stellar mass selection. It is evident that after the mass segregation, we are able to compare similar -- in a statistical manner -- samples of galaxies. In the same table we also present the duration of each redshift period (in Gyr), and the number of LTGs and ETGs, as well as dynamically young and old groups found in each bin.

\begin{figure*}
\begin{center}
\includegraphics[scale=0.9]{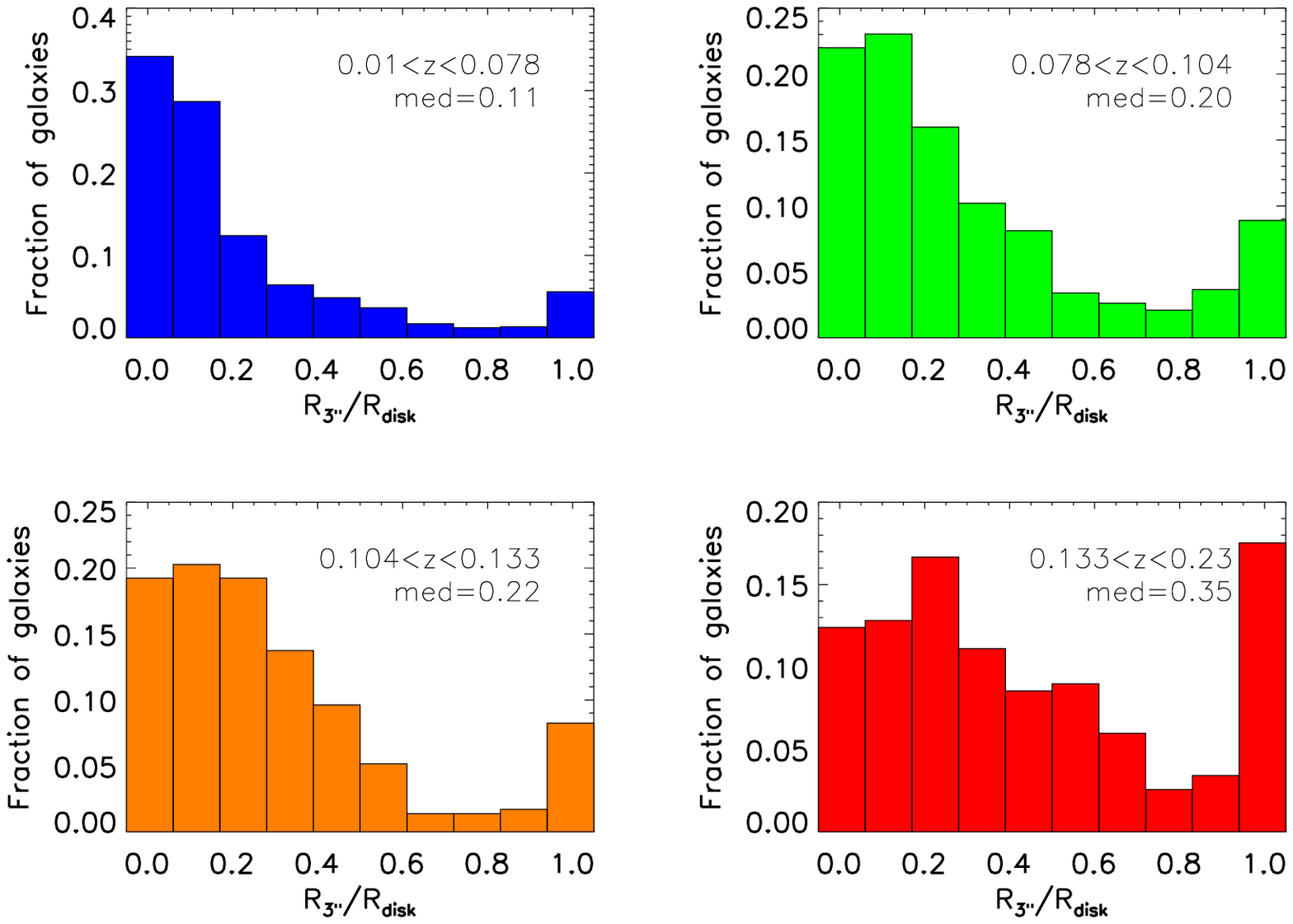}
\caption{Histograms of the ratios of the $SDSS$ 3\arcsec spectro-fiber over the disk radius for the individual disk-galaxies in our sample, at the different redshift bins (colour-coding as in Fig.~\ref{fig1}). In the upper right corners we indicate the redshift ranges and the median values for each bin. (A coloured version of this figure is available in the online journal) }
\label{fig_cont}
\end{center}
\end{figure*}

\begin{table*}
\begin{minipage}{180mm}
\caption{Number of galaxies and median stellar masses in the different redshift bins.}
\label{tab_bin}
\begin{center}
\begin{tabular}{cccccccc}
\hline \hline
z bin - range  & No. of galaxies$^{a}$ & log(M$_{star}$) [M$_{\odot}$]$^{a}$ & No. of galaxies$^{b}$ & log(M$_{star}$) [M$_{\odot}$]$^{b}$ & T$^{c}$ [Gyr] & LTGs/ETGs$^{d}$ & DY/DO$^{e}$ \\
\hline
Bin1 0.010-0.078  &    2,237 &   10.26$\pm$0.20 &    1,109 & 10.75$\pm$0.32  & 1.0 & 442/667 & 61/216 \\
\rowcolor{Gray} Bin2 0.078-0.104  & 1,717 & 10.66$\pm$0.25 & 1,119 & 10.80$\pm$0.32  & 0.4 & 359/760 & 50/230 \\
Bin3 0.104-0.133  & 1,444 & 10.86$\pm$0.29 & 1,111 & 10.85$\pm$0.33  & 0.4 & 330/780 & 34/244 \\
\rowcolor{Gray} Bin4 0.133-0.230  &    1,489 & 11.09$\pm$0.28 &    1,016 & 10.97$\pm$0.34  & 1.0 & 328/687 & 38/217 \\
\hline
All redshifts  & 7,417 & 10.76$\pm$0.12 & 4,355 & 10.84$\pm$0.18  & 2.8 & 1,459/2,894 & 183/907 \\
\hline
\end{tabular}
\end{center}
$^{a}$ Total number of galaxies in each bin.\\
$^{b}$ Number of galaxies in each bin with stellar masses of 10.5$\le$log(M$_{star}$)$\le$11.2 M$_{\odot}$.\\
$^{c}$ Duration of each redshift period, according to $\Lambda$CDM cosmology.\\
$^{d}$ Number of late-type (LTGs) and early-type (ETGs) galaxies in each bin, after the mass selection.\\
$^{e}$ Number of dynamically young (DY) and old (DO) groups in each bin, after the mass selection.\\
\end{minipage}
\end{table*}%

In \S3.1, we classified the morphologies of our galaxies based on the fitted radial profiles and the selection criteria described in \citet{Simard11}. Nevertheless, due to their larger distances, LTGs at higher-z's might appear more ``spheroidal'' with result to be misclassified as ETGs. In that case, we should expect to observe more ETGs as we are probing at higher-z. To examine this, we have applied -- in addition to our current -- the classifications proposed by \citet{Balogh04} and \citet{Tanaka05}. These authors classified galaxies into star-forming (thus, LTG) or quiescent (thus, ETG), based on their H$\alpha$ equivalent widths (EW(H$\alpha$)), with star-forming galaxies having EW(H$\alpha$)$>$4\AA{}. Comparing the results of these classifications with the one we performed in \S3.1, we find differences of $<$3\% at the various redshift bins. In addition, using the published galaxy morphologies from the ``Galaxy Zoo''\footnote{Available at http://www.galaxyzoo.org} \citep{Lintott11}, we found more than 95\% agreement, for classifications with $>$80\% confidence, suggesting that the fraction of misclassified galaxies in our sample is likely very low. 

Finally, one more probable bias associated with our data, could emerge from the fact that as we are probing at higher redshifts, the nuclear spectra will be increasingly contaminated by their galaxy-hosts (due to the finite size of the slit/fiber). Using the results of the fitted radial profiles of our galaxies, from \citet{Simard11}, we plot in Fig.~\ref{fig_cont} the distributions of the ratios of the 3$''$ SDSS spectro-fiber over the disk radii, at the different redshift bins. Our results show that at z$<$0.104 this contamination is insignificant, however at 0.104$<$z$<$0.133 around 10-15\%, and at 0.133$<$z$<$0.230 more than 20\% of the disk galaxies may suffer a significant contamination in their nuclear spectra. \citet{Moran02} showed that galaxy nuclear spectra suffering a contamination from their disks would move towards the lower-left portion of the classic BPT diagram, due to the dilution of their spectral lines, therefore often misclassifying AGN as SFN. However, a more recent study by \citet{Maragkoudakis14} revealed that galaxies with extra-nuclear star formation can show higher [O{\sevensize III}]$\lambda5007$/H$\beta$ line ratios, since lower metallicity H{\sevensize II} regions in the outer parts of galaxy discs are also capable of producing high-excitation emission lines, implying that the results of such a contamination might rather be challenging to interpret.

\section{Results}

\subsection{Nuclear activity in late-type galaxies}
\begin{figure}
\begin{center}
\includegraphics[scale=0.52]{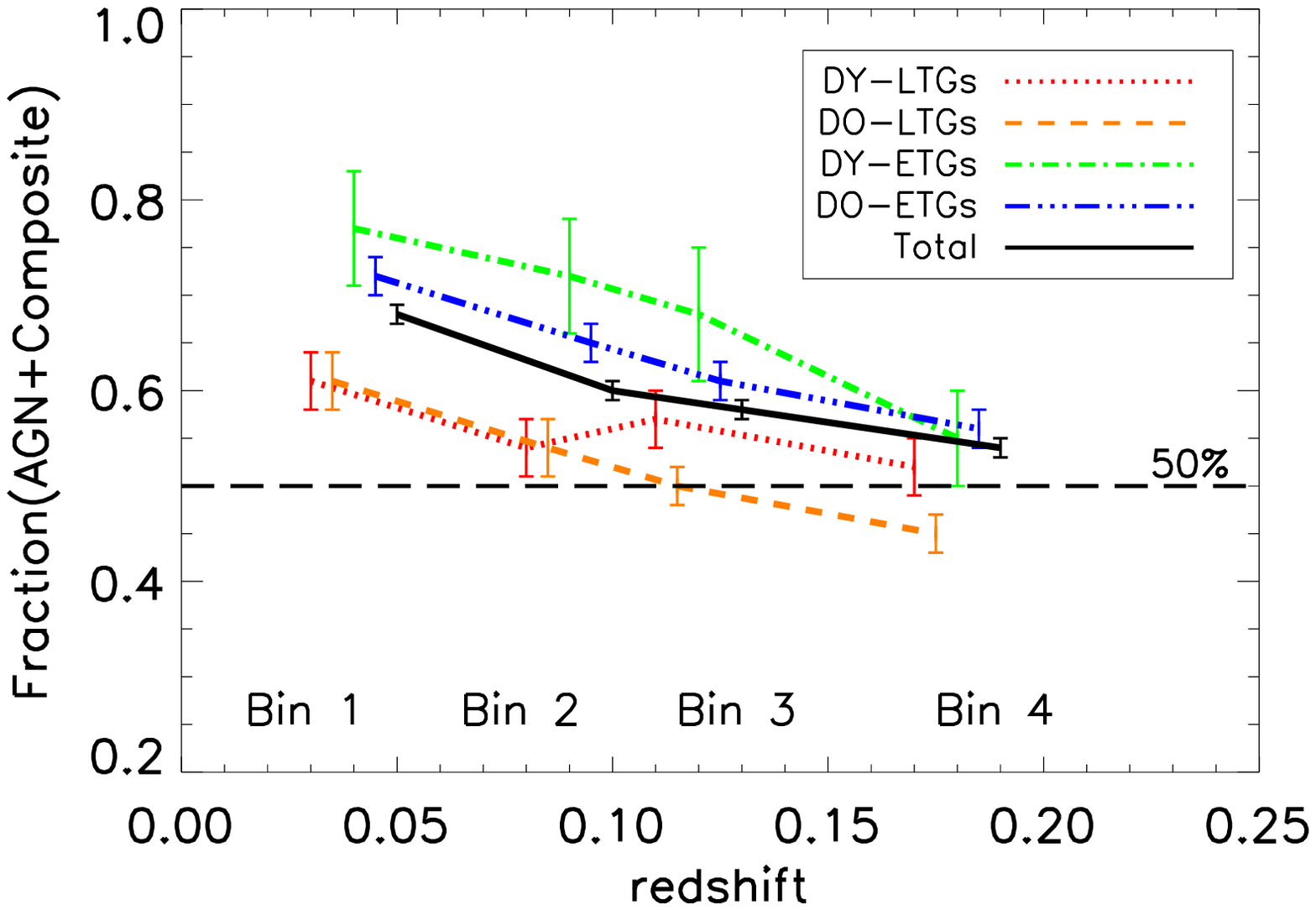}
\caption{Fractions of the AGN+composite hosting galaxies over the total population of galaxies in each redshift bin, as a function of redshift, for the four different sub-samples we presented in Table~\ref{tab1} (DY-LTGs in red, DO-LTGs in orange, DY-ETGs in green and DO-ETGs in blue lines), as well as for the total sample (in black solid line). X-axis values are taken in the middle of each redshift bin, slightly displaced to resolve better. Uncertainties denote 1$\sigma$ Poissonian errors. (A coloured version of this figure is available in the online journal)}
\label{fig45}
\end{center}
\end{figure}

As presented in the previous section, to study the evolution of the AGN activity in our sample, we have separated it into four different redshift bins. We used galaxies within the mass-range of 10.4$\le$log(M$_{star}$)$\le$11.3 M$_{\odot}$, as described in \S3.4. The corresponding fractions of the spectroscopic classifications for the different subsamples are presented in Table~\ref{tab1}. In the discussion that follows, we will refer as AGN hosts, to all galaxies classified either as AGN or composite objects. In Fig.~\ref{fig45}, we present how the AGN fractions\footnote{As AGN fraction, we define the fraction of AGN-hosting galaxies in our sample (it is different from the $f_{AGN}$ mentioned in \S3.3)} of the various sub-samples change with redshift. We notice that the number of  AGN-hosting galaxies in our sample (marked with black solid line) increases by $\sim$15\% moving towards lower-z's. When we separate them according to the dynamical state of their group (also presented in the same table), the AGN fractions of the LTGs found in dynamically old groups increase from 45\% to 61\%. Moreover, their Sy2 fractions, which is 7-8\% beyond z$\sim$0.133, is changing to 14\%-28\% at lower redshifts (the range depends on the classification method; see \S3.2). The above results suggest a significant increase of the AGN-hosting galaxies over the past 3 Gyr, accompanied by the corresponding increase of their Sy2-hosts fraction. On the other hand, LINER fractions do not seem to change at all. In contrast, the AGN activities of the LTGs in dynamically young groups, do not seem to change during the same period. Despite their AGN fraction seems to slightly increase (within the 1-$\sigma$ level), their Sy2 fraction remains almost constant. In Table~\ref{tab_yair}, we present the corresponding fractions of the local isolated field and interacting pair LTGs samples (described in \S2.2), found in the same mass range with our galaxies. Interestingly, we can notice that there are no statistical differences between the different environments in the local Universe, with all of them having AGN-hosting galaxy fractions of about 60\%. However, we have to stress here that due to the uncertainties in some of the comparison samples, the results of such a comparison are not conclusive.

\begin{table*}
\begin{minipage}{140mm}
\caption{Fractions of the type of nuclear activity in our sample, found in galaxies of 10.4$\le$log(M$_{star}$)$\le$11.3 M$_{\odot}$.}
\label{tab1}
\begin{center}
\begin{tabular}{ccccccc}
\hline \hline
redshift bin  & SFN$^{a}$ & Composite$^{a}$ & AGN$^{a}$ & AGN+Comp.$^{a}$ & LINER$^{b}$ & Sy2$^{b}$ \\
\hline
\multicolumn{7}{c}{Late-type galaxies in dynamically young groups (DY-LTGs)}\\
\hline
Bin1  & 39$\pm$4\% & 31$\pm$4\% & 30$\pm$4\% & 61$\pm$3\% & 13-37$\pm$4\% & 16-37$\pm$5\% \\
\rowcolor{Gray} Bin2  & 46$\pm$6\% & 33$\pm$5\% & 21$\pm$3\% & 54$\pm$3\% & 20-22$\pm$6\% & 14-18$\pm$5\% \\
Bin3  & 43$\pm$6\% & 39$\pm$6\% & 19$\pm$4\% & 57$\pm$3\% & 25$\pm$7\% & 8-18$\pm$4\% \\
\rowcolor{Gray} Bin4  & 48$\pm$7\% & 34$\pm$6\% & 18$\pm$5\% & 52$\pm$3\% & 15-24$\pm$6\% & 12-21$\pm$6\% \\
\hline
\multicolumn{7}{c}{Late-type galaxies in dynamically old groups (DO-LTGs)}\\
\hline
Bin1  & 39$\pm$3\% & 32$\pm$3\% & 29$\pm$3\% & 61$\pm$3\% & 21-26$\pm$4\% & 14-28$\pm$4\% \\
\rowcolor{Gray} Bin2  & 46$\pm$4\% & 35$\pm$4\% & 19$\pm$3\% & 54$\pm$3\% & 23-29$\pm$4\% & 9-25$\pm$6\% \\
Bin3  & 50$\pm$4\% & 29$\pm$3\% & 21$\pm$2\% & 50$\pm$2\% & 16-26$\pm$4\% & 24-32$\pm$6\% \\
\rowcolor{Gray} Bin4  & 55$\pm$5\% & 30$\pm$4\% & 14$\pm$3\% & 45$\pm$2\% & 21-29$\pm$7\% & 7-8$\pm$4\% \\
\hline
\multicolumn{7}{c}{Early-type galaxies in dynamically young groups (DY-ETGs)}\\
\hline
Bin1  & 23$\pm$7\% & 23$\pm$7\% & 54$\pm$11\% & 77$\pm$6\% & 41-52$\pm$12\% & 15-37$\pm$7\% \\
\rowcolor{Gray} Bin2  & 28$\pm$7\% & 25$\pm$7\% & 47$\pm$10\% & 72$\pm$6\% & 35$\pm$11\% & 15-23$\pm$7\% \\
Bin3  & 32$\pm$11\% & 37$\pm$13\% & 32$\pm$11\% & 68$\pm$7\% & 23-31$\pm$13\% & 8-23$\pm$7\% \\
\rowcolor{Gray} Bin4  & 45$\pm$13\% & 41$\pm$13\% & 14$\pm$7\% & 55$\pm$5\% & 8-22$\pm$8\% & 0-8$\pm$8\% \\
\hline
\multicolumn{7}{c}{Early-type galaxies in dynamically old groups (DO-ETGs)}\\
\hline
Bin1  & 28$\pm$1\% & 24$\pm$1\% & 48$\pm$2\% & 72$\pm$2\% & 33-34$\pm$3\% & 13-30$\pm$2\% \\
\rowcolor{Gray} Bin2  & 35$\pm$2\% & 26$\pm$1\% & 39$\pm$2\% & 65$\pm$2\% & 34-38$\pm$3\% & 15-18$\pm$2\% \\
Bin3  & 39$\pm$2\% & 26$\pm$1\% & 35$\pm$1\% & 61$\pm$2\% & 30-31$\pm$3\% & 13-18$\pm$2\% \\
\rowcolor{Gray} Bin4  & 44$\pm$2\% & 26$\pm$2\% & 31$\pm$2\% & 56$\pm$2\% & 26-30$\pm$4\% & 15-22$\pm$3\% \\
\hline
\end{tabular}
\end{center}
$^{a}$Classifications from the classic BPT-diagram, using the [OIII]$\lambda5007$/H$\beta$ vs the [NII]$\lambda6584$/H$\alpha$ line ratios.\\
$^{b}$Classifications based on the \citet{Kewley06}, using both the [SII]$(\lambda6717+6731)$/H$\alpha$ and [OI]$\lambda6300$/H$\alpha$ lines, for galaxies already classified as AGN or Composite from the classic BPT diagram.\\
Uncertainties denote the 1$\sigma$ Poissonian errors.\\
\end{minipage}
\end{table*}%

\begin{table*}
\begin{minipage}{140mm}
\caption{The nuclear activity of the comparison samples.}
\label{tab_yair}
\begin{center}
\begin{tabular}{cccc}
\hline \hline
Sample  & SFN & AGN & AGN+Comp. \\
\hline
\multicolumn{4}{c}{late-type galaxies}\\
\hline
\rowcolor{Gray} Isolated field$^{a}$  & 41$\pm$6\% & 10$\pm$3\% & 59$\pm$7\% \\
S+S pairs$^{b}$       & 41$\pm$7\% & 34$\pm$6\% & 59$\pm$8\% \\
\rowcolor{Gray} S+E pairs$^{b}$       & 42$\pm$13\% & 38$\pm$13\% & 58$\pm$16\% \\
\hline
\multicolumn{4}{c}{early-type galaxies}\\
\hline
\rowcolor{Gray} Isolated field$^{a}$  & 7$\pm$7\% & 93$\pm$26\% & 93$\pm$26\% \\
S+E pairs$^{b}$      & 5$\pm$5\% & 85$\pm$20\% & 95$\pm$21\% \\
\rowcolor{Gray} E+E pairs$^{b}$      & 6$\pm$6\% & 94$\pm$22\% & 94$\pm$22\% \\
\hline
\end{tabular}
\end{center}
$^{a}$Isolated field galaxies from \citet{Hernandez13}, with 10.4$\le$log(M$_{star}$)$\le$11.3 M$_{\odot}$.\\
$^{b}$Isolated interacting pair galaxies from \citet{Hernandez14}, with 10.4$\le$log(M$_{star}$)$\le$11.3 M$_{\odot}$ (Following these authors' symbolism, S denotes the LTGs, and E the ETGs).\\
These fractions correspond to the 87\% and 82\% of the field and interacting pair galaxy samples, respectively, which were detected to have emission lines.\\ 
Uncertainties denote the 1$\sigma$ Poissonian errors.\\
\end{minipage}
\end{table*}%

In the first columns of Tables~\ref{tab1} and \ref{tab_yair}, we also present the number of SFN galaxies in our sample, as well as the comparison samples. Again, we find no differences between the field isolated galaxies (41\%), the S+S and S+E pairs (with 41\% and 42\%, respectively), and the dynamically young and old groups (both having 39\% SFN at low-z's). On the other hand, looking at higher-z's, we can see that both dynamically young and old LTGs, have similar SFN (of 48\% and 55\%, respectively). These results are also consistent with our observations of their UV-optical colours. Studying the [NUV-r] colours of the dynamically old LTGs we notice significant reduction, of almost 40\%, of the blue star forming galaxies (blue cloud) at lower redshifts, and at the same time an increase of 30\% and 10\% of the green valley and red sequence galaxies (quiescent galaxies), respectively (Bitsakis et al. - in prep.). On the contrary, this blue cloud decrease is not observed to be significant for the LTGs in dynamically young groups (only 15\%).   

\begin{table*}
\begin{minipage}{120mm}
\begin{center}
\caption{Contribution of the AGN to the total IR luminosity ($f_{AGN}$) of the galaxies in the mass range 10.4$\le$log(M$_{star}$)$\le$11.3 M$_{\odot}$.}
\label{tab2}
\begin{tabular}{crrrr}
\hline \hline
redshift bin  & $f_{AGN}$ & $f_{AGN}$ & $f_{AGN}$ & $f_{AGN}$ \\
  & DY-LTGs & DO-LTGs & DY-ETGs & DO-ETGs \\
\hline
Bin1  &   3$\pm$1\% &   4$\pm$1\% &   7$\pm$1\% &   6$\pm$1\% \\
\rowcolor{Gray} Bin2  &   5$\pm$1\% &   5$\pm$1\% &   10$\pm$1\% &   10$\pm$1\% \\
Bin3  &   6$\pm$1\% &   8$\pm$1\% & 10$\pm$1\% & 14$\pm$1\% \\
\rowcolor{Gray} Bin4  & 10$\pm$1\% & 11$\pm$1\% & 16$\pm$1\% & 17$\pm$1\% \\
\hline
\end{tabular}
\end{center}
With DY and DO we denote the dynamically young and old groups, respectively. \\
Note that for $f_{AGN}$$<$5\% the existence of an AGN is highly uncertain (see \S3.3).\\
Uncertainties denote the 1$\sigma$ Poissonian errors.
\end{minipage}
\end{table*}%

In a first attempt to interpret these results we rely on the conclusions of \citet{Bitsakis11}, for HCG galaxies. According to these, as the time progresses the multiple encounters each compact group galaxy experiences, will result in the built-up of its stellar mass, as well as to the depletion (via star formation and/or tidal stripping) of a significant fraction of its gas (in dynamically old groups is more than 70\%; derived for HCG galaxies, see \citealt{Verdes01,Borthakur10}) and dust content \citep[more than an order of magnitude lower from what is observed in dynamically young and isolated field LTGs; see][]{Bitsakis14}. In dynamically old groups, this effect is expected to appear stronger, since they are denser, more dynamically evolved, and their galaxies have experienced most probably a higher rate of tidal encounters. The high AGN fractions at lower redshifts can be explained assuming that, as the star formation of the compact group LTGs slowly fades-out (due to the depletion of gas, mentioned earlier), it reveals a weaker AGN activity, might also been triggered by the dynamical interactions. This assumption is consistent with the scenario proposed by \citet{Krongold02}, where galaxy interactions/merging initially trigger central starbursts and subsequently reveal a Sy2 and eventually a Sy1 nucleus (thus increasing the fraction of AGN hosting galaxies). To explain the transition between Sy2 and Sy1 nuclei, the following scenarios have been proposed:  a very powerful, fully-obscured Sy1, which due to feedback will either completely wipe-out the torus or make it clumpy \citep{Sirocky08}, or a sequence on the AGN power that depends either on the low accretion rates or the small mass of the SMBH \citep[e.g.][]{Nicastro00}. \citet{Krongold02} suggested that transitions from Sy2 to Sy1 AGN may have been delayed as much as 1 Gyr (upper limit). During this period, compact group galaxies will, most probably, experience a large number of encounters \citep[having dynamical times of $\sim$100 Myr;][]{Hickson97}, which will result in the depletion of large amounts of their interstellar gas (via induced star formation and/or tidal stripping), thus preventing  high accretion rates towards the SMBH, required to form a BLR and thus a potential Sy1 nuclei. This conclusion is consistent with the results of \citet{Nicastro03}, where the absence of HBLRs in Sy2 AGN can be regulated by the accretion rate of the SMBH.

\begin{table*}
\begin{minipage}{120mm}
\begin{center}
\caption{Median AGN H$\alpha$ luminosities}
\label{tab3}
\begin{tabular}{ccccc}
\hline \hline
redshift bin  & DY-LTGs & DO-LTGs & DY-ETGs & DO-ETGs \\
          & $\times$10$^{40}$erg s$^{-1}$ & $\times$10$^{40}$erg s$^{-1}$ & $\times$10$^{40}$erg s$^{-1}$ & $\times$10$^{40}$erg s$^{-1}$ \\
\hline
Bin1  & 0.67$\pm$0.07 & 0.58$\pm$0.05 & 0.26$\pm$0.05 & 0.34$\pm$0.02 \\
\rowcolor{Gray} Bin2  & 1.63$\pm$0.23 & 0.81$\pm$0.10 & 0.57$\pm$0.11 & 0.48$\pm$0.03 \\
Bin3  & 1.64$\pm$0.25 & 2.07$\pm$0.25 & 0.76$\pm$0.21 & 0.54$\pm$0.03 \\
\rowcolor{Gray} Bin4  & 3.70$\pm$0.63 & 2.41$\pm$0.37 & 1.51$\pm$0.43 & 1.03$\pm$0.07 \\
\hline
\end{tabular}
\end{center}
With DY and DO we denote the dynamically young and old groups, respectively. Uncertainties denote the 1$\sigma$ Poissonian errors.
\end{minipage}
\end{table*}%

The above scenario predicts the existence of only a very small fraction of Sy1 nuclei in compact groups. Using the classification described in \S3.1, we find that only $<$1\%, of the galaxies in our sample, may have broad-lines. The almost complete absence of such objects does not support the inclination scenario, at least in most cases, according to which we should be able to observe more than 20\% of them, in the local Universe \citep{Ho97}. It is possible, though, that heavy dust obscuration could hide them from us. Using, the advent of infrared extragalactic surveys, such as the {\it WISE} All Sky Release, we are able to unveil the mystery of the emission from dust obscured regions. Based on the mid-IR colour selection proposed by \citet[][see Fig.~\ref{fig5}]{Mateos12} we find that only 0.5\% of the galaxies in our sample could be dust obscured AGN. In addition, according to the WISE [3.4]-[4.6]$>$0.8 (Vega mags) colour selection, by \citet{Stern12}, this fraction is also 0.3\%. Therefore, it is evident that obscuration is not a factor that affects the absence of Sy1 sources from our sample. 

An alternative method to detect IR luminous AGN has been presented in \S3.3 \citep[and also described in more detail in][]{Ciesla14}. There, we showed how our SED modeling is able to estimate the fraction of the total infrared luminosity of a galaxy, due to an AGN ($f_{AGN}$). In Table~\ref{tab2}, we present the results of the fitting, separating galaxies according to their optical morphologies, the dynamical state of their group, and their redshift. Firstly, we can see the nearly complete absence of AGN activity in the local compact group galaxies, independent of morphology and group state (having $f_{AGN}$$<$5\%). On the other hand, it is also evident the rise of 10\% that occurs at redshifts beyond 0.133. We have to remind here to the reader, that according to \citet{Ciesla14} CIGALE's results can be reliable only for $f_{AGN}$$>$10\%. Between 5\%$<$$f_{AGN}$$<$10\%, the uncertainties are very high, and  below 5\%, its impossible to identify an AGN. 

\citet{Martinez10} showed that even though 45\% of HCG galaxies in their sample, host an AGN, the majority appear to have low-luminosities (known as low-luminosity AGN; LLAGN), with a median H$\alpha$ luminosity of 0.71$\times$10$^{40}$ erg sec$^{-1}$. We estimate the corresponding value for the AGN hosting galaxies in our sample to be 0.76$\pm$0.01$\times$10$^{40}$ erg sec$^{-1}$, which is in excellent agreement with their result. In Table~\ref{tab3}, we also present the corresponding H$\alpha$ luminosities of the galaxies in our sample, separating them by their morphologies and evolutional state of their group. We can see that there is no statistical difference between the various sub-samples of LTGs and ETGs at the different redshifts, but there is a significant increase (more than an order of magnitude), that occurs at higher redshifts, which is observed in all the sub-samples. On the other hand, local interacting pairs and isolated galaxies display luminosities of $\sim$1-2$\times$10$^{40}$ erg sec$^{-1}$, which are similar to the corresponding compact group luminosities at higher redshifts. The fact that we observe a significant decrease in the AGN H$\alpha$ luminosities and the f$_{AGN}$, may suggest that the AGN activity in groups became weaker during the last 3 Gyr, possibly as a consequence of the gas starvation scenario, initially proposed by \citet{Martinez10}.  

\subsection{Nuclear activity in early-type galaxies}
In Table~\ref{tab1} and Fig.~\ref{fig45}, we have also presented the corresponding AGN and SFN fractions of the ETGs in our compact group sample. Similarly to what we have seen for LTGs, the number of AGN-hosting ETGs found in dynamically old groups is increasing towards lower redshifts, by 16\%. Similarly, the AGN fraction of ETGs found in dynamically young groups seem also to increase by 15\%, in the 2-$\sigma$ level. Examining the Sy2's in both sub-samples, it appears that in contrast with the dynamically young groups, where an increase of 10-20\% is observed, in old groups these fractions remain high but constant. Similar is the evolution of LINERs with the fractions of the first increasing, and those of the latter remaining almost constant and much higher than those in LTGs. Comparing our findings with those of the control samples (see bottom panel of Table~\ref{tab_yair}) we notice that, emission line isolated field galaxies have a corresponding fraction of AGN hosts of 93\%, and E+S and E+E pairs have 95\% and 94\%, respectively. Unfortunately, due to the very large uncertainties, we cannot derive statistically significant conclusions from these comparisons. 

In an effort to interpret the observed differences between the Sy2 and LINER fractions of the AGN-hosting ETGs in our sample, we propose the following possibilities: in dynamically young groups we may observe the morphological transformation of Sa's and S0's into ellipticals. As discussed in \S4.1, the remaining star formation activity of Sa's and S0's is expected to fade-out and subsequently reveal an AGN nucleus (which most probably is a Sy2). During that period, S0's already at later phases of evolution may eventually evolve into ellipticals (usually having LINERs; \citealt{Ho97b}). On the other hand, by definition dynamically old groups host large numbers of ETGs at all epochs, which are expected to be either evolved S0's or ellipticals \citep[see][]{Hickson82, Bitsakis11}. Therefore, both Sy2 as well as LINER fractions are expected to be high at all times, but also constant. However, we should mention once more that Sy2 and LINER fractions discussed here are based on the AGN definition of the classic BPT diagram, while the true nature of LINERs is still uncertain. More specifically a number of studies \citep[e.g.][]{Yan12} suggest that the excitation mechanisms behind LINERs are a combination of post-AGB stars and shocks. \citet{Singh13} argued that post-AGB stars are ubiquitous and their ionising effects would be present in all galaxies, having stellar populations older than 1 Gyr, unless they are out-shined by a much brighter young stellar population and/or an AGN. On the other hand, \citet{Gonzalez06} showed that at least 60\% of the LINER galaxies in their sample could host an AGN, based on their X-ray morphologies and spectra. In addition, more recently \citet{Dopita15} also presented the case of the LINER elliptical NGC1052, where an accreting SMBH was detected.

Unfortunately, due to the morphological classification we applied in our sample, we cannot distinguish between elliptical and S0 galaxies, in order to verify the proposed scenario. Nevertheless, examining their bulge-to-disk ratios (B/T),  we can see their tendency to be more ``disky'' (having lower B/T ratios, as derived from \citealt{Simard11}) or ``bulgy'' (with higher B/T ratios). We find that in local dynamically old groups, ETGs have a median B/T=0.69$\pm$0.07 that decreases to 0.62$\pm$0.06 at z$>$0.133. In dynamically young groups the corresponding B/T ratios change from 0.64$\pm$0.18 (at Bin1) to 0.52$\pm$0.25 (at z$>$0.133), suggesting that they are not significant differences in the morphologies. 

Even though the above scenarios can explain the observed nuclear activities for a large number of ETGs in our sample, there might always be the possibility that some of them have increased their gas content, via accretion/merging from gas-rich companions, which are more common in dynamically young groups. Indeed, in \citet{Bitsakis14}, we showed that about 25\% of the Hickson compact group ETGs display blue colours and enhanced star formation activities. Examining their dust masses and star formation rates, we found that they are almost an order of magnitude higher in respect to those of field ETGs, suggesting that compact group ETGs may have enhanced their gas and dust content via accretion from gas rich members and/or the merging of dwarf companions.

\begin{figure}
\begin{center}
\includegraphics[scale=0.51]{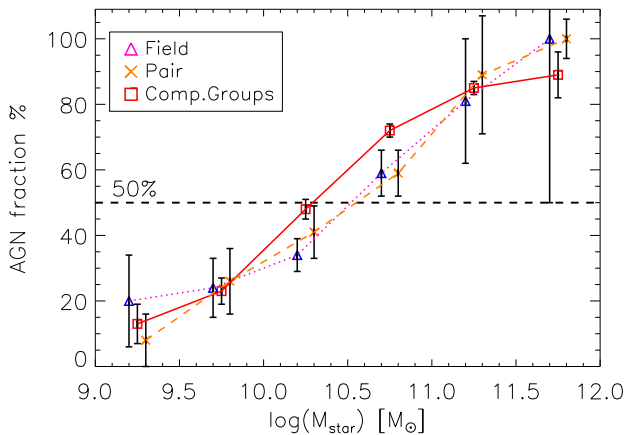}
\caption{AGN fractions of the compact group (in red squares and continuous line), isolated field (in blue triangles and dotted line) and pair (in orange X's and dashed line) galaxies, as a function of their stellar mass. For the purposes of comparison with the control samples we have selected CG galaxies at z$<$0.05. X-axis values are taken in the middle of each stellar mass bin, slightly displaced to resolve better. The 50\% dashed line was placed to show where AGN rules-over SFN activity. Error-bars denote the 1$\sigma$ Poissonian errors. (A coloured version of this figure is available in the online journal)}
\label{fig6}
\end{center}
\end{figure}

Finally, examining the AGN H$\alpha$ luminosities of the dynamically young and old ETG samples (see Table~\ref{tab3}), we can see them both decreasing more than an order of magnitude, as we are moving towards lower-z's. As it was suggested for compact group LTGs, in \S4.1, this could be an indirect indication of the gas depletion these sources had experienced, throughout time. Moreover, we can see that dynamically old AGN-hosting ETGs display lower H$\alpha$ luminosities, at all redshifts, with respect to those of the dynamically young ETGs. These results imply that ETGs in dynamically young groups may have larger gas contents, either due to accretion from gas rich neighbours, or because they mostly host later-type ETGs, in the Hubble classification scheme. 

\subsection{The relation between environment and AGN activity at lower mass galaxies}

It is generally accepted that the nuclear activity of massive galaxies is mostly dominated by AGN rather than circumnuclear star formation \citep[i.e.][and references therein]{Caputi14}. \citet{Hernandez13, Hernandez14} showed this tendency in isolated field and interacting pair galaxies. In figures 6 and 4 of these publications, respectively, one can notice that AGN in isolated field galaxies start dominating the emission with respect to SFN, at log(M$_{star}$)$\ge$10.7 M$_{\odot}$, whereas in pairs this occurs at log(M$_{star}$)$\ge$10.5 M$_{\odot}$. In addition, \citet{Pimbblet13} showed that AGN in clusters are mostly found in galaxies with log(M$_{star}$)$\ge$10.7 M$_{\odot}$. In Fig.~\ref{fig6}, we present the AGN fractions as a function of the stellar mass, for the  galaxies in our sample (presented with the red squares and the solid line), having z$<$0.05 in order to compare them with those of the isolated field and interacting pair samples (shown with blue triangles and the dotted line, and orange X's and the dashed line, respectively). The three distributions do not appear to have any statistical difference, due to their large uncertainties, yet one can see the tendency of the compact group galaxies to exceed the 50\% line (where AGN rule-over SFN), at lower stellar masses, than the comparison samples. 
 
\begin{figure}
\begin{center}
\includegraphics[scale=0.51]{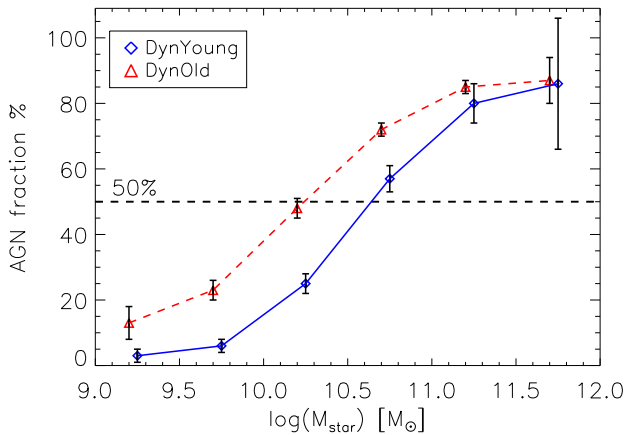}
\caption{AGN fractions as a function of stellar mass of the dynamically young (in blue diamonds and continuous line) and dynamically old (in red triangles and dashed line) galaxies in our sample. Explanations as in Fig.~\ref{fig6}. (A coloured version of this figure is available in the online journal)}
\label{fig7}
\end{center}
\end{figure}

In Fig.~\ref{fig7}, we present the same figure, this time having all compact groups found at all redshifts. We also separate CG galaxies according to the dynamical state of their group (into dynamically young, blue solid line, and old, red dashed line). It is evident that the incidence of the AGN activity in dynamically young groups starts prevailing over SFN at masses log(M$_{star}$)$\ge$10.6 M$_{\odot}$ (similar to those of the isolated field and interacting pair galaxies), whereas in dynamically old groups at log(M$_{star}$)$\ge$10.2 M$_{\odot}$. We perform a Kolmogorov-Smirnov (KS) analysis, to test the hypothesis the two distributions being drawn from the same parent distribution. The KS probability is P$_{KS}$=0.2\%, thus rejecting the null hypothesis, and suggesting they are different, with a level of significance of more than 99.9\%. These results suggest that the more dynamically evolved the group, the lower the stellar mass at which AGN dominate the star formation activity of their members. 

To interpret these results, we considered the following scenarios: $(i)$ either interactions enhance AGN activity, or $(ii)$ star formation fades-out faster during the evolution of these galaxies. Nevertheless, as we already presented there {\bf is} no evidence to support an enhancement of the AGN activity due to interactions. On the contrary, we showed how multiple encounters result to the decrease of the AGN luminosities. However, gas consumption, due to tidally induced star formation \citep{Fedotov11}, tidal stripping \citep[i.e.][]{Verdes01}, and shock excitation of the interstellar medium \citep{Cluver13, Alatalo14}, will result to the depletion of the {\bf cold} gas reservoir of these galaxies or will make it unable to form stars (in the case of shocks). These results are also consistent with the observed absence of Sy1 nuclei in our sample, where the depletion of gas may led to small accretion rates. From the above it is evident that the the prevalence of AGN activity over star formation, is connected with the evolutional state of the environment (or the  number of encounters galaxies had experienced).

\section{Discussion: The role of compact groups in the AGN zoo}
The results of this study suggest that compact groups can influence the evolution of the nuclear activity of their galaxies in an indirect manner, by affecting the available amount of gas. How does this result can be compared to the picture we already have about AGN? Do compact groups really differentiate their nuclear properties from galaxies in other environments? It seems that our results do not support the AGN unification scheme, according to which the only differentiation between Sy1 and Sy2 sources is caused by the dust obscuration of the torus. The almost complete absence of Sy1 nuclei in compact groups, in addition to the low AGN H$\alpha$ luminosities, suggest that the decrease of the gas reservoir may lead to low accretion rates, which according to \citet{Nicastro00} and \citet{Elitzur06} have as result to prevent from existence powerful Sy1 nuclei. The observed absence of such sources (also according to mid-IR colour diagnostics and SED modeling) is also consistent with what it was observed in pairs of galaxies \citep{Hernandez14}. Both results agree with the findings of \citet{Dultzin99} and more recently \citet{Koulouridis06} and \citet{Villarroel14}, where galaxies with close companions (found in distances closer than $\sim$150 kpc; in CGs almost all galaxies are found in smaller distances, see \citealt{Bitsakis11}) are more likely to host a Sy2 nucleus, rather than a Sy1. 

As we mentioned in \S4.1 the above results led \citet{Krongold02}, and more recently \citet{Koulouridis14}, to propose an evolutionary sequence in the nuclear activity. Initially interactions may cause the collapse of gas towards the inner regions of the galaxies, which will trigger circumnuclear star formation, as well as a weak, dust enshrouded AGN, which may appear as a HBLR Sy2 nucleus. Then, the ever increasing presence of AGN feedback will lead to its transformation into a Sy1, and, eventually, the consumption and/or loss of its fuel might transform it back into a non-HBLR Sy2.  However, the results of our current work, as well as previous AGN compact group studies \citep[i.e.][]{Martinez10, Sohn13}, suggested that most compact group galaxies do not seem to complete that process. The absence of Sy1 nuclei and the low AGN H$\alpha$ luminosities, which are reducing towards higher redshifts, imply that this evolutionary sequence stops during its early phase. In \S4.1, we mentioned that in typical time scales, where a Sy2 can turn into a Sy1, galaxies in compact groups will most probably experience a large number of encounters, which will, eventually, prevent this transformation. Galaxy mergers also do not seem to be a possible alternative to enrich galaxies and fuel a powerful AGN accretion. It has been shown that, most such events in CGs take place under dry conditions, thus no gas enrichment of the AGN-host can occur \citep{Coziol07, Konstantopoulos10}.

Another result of this work is that the number of compact group AGN-hosts (especially those of galaxies found in dynamically old groups) increases at lower redshifts. At a first glance, these results seem to be inconsistent with recent observations, where the AGN number densities have decreased significantly since z=1-2 \citep[e.g.][]{Barger05}. Nevertheless, upon careful examination we find that it is not the number of compact group AGN-hosts that increases, but the fraction of SFN that decreases due to the depletion of gas, which affects star formation. These results are also confirmed by Bitsakis et al. - in prep., where the star formation activity of the late-type galaxies is significantly decreasing since z$\sim$0.2. Moreover, in a number of compact group studies it is already presented how the star formation activity of compact group late-type galaxies can be seriously affected by the depletion of gas (i.e. the specific star formation rates of dynamically old group LTGs are more than an order of magnitude lower to those of similar galaxies in the field; see \citealt{Bitsakis11} and references therein). The past interaction-triggered star formation activity, the tidal stripping -- ram pressure effects have been shown to be weak in HCGs \citep{Rasmussen08}, even though \citet{Desjardins14} still advocate their influence -- and the shock excitation of their remaining interstellar medium \citep{Cluver13, Alatalo14}, resulted either in the loss of their gas reservoir or its inability (when shocked) to form stars. These effects are expected to influence stronger the star formation activity that also occurrs in their disks, thus making them more vulnerable to gravitational interactions (disk material can be stripped easier). Moreover, according to the AGN ``downsizing'' scenario, there is an observed decrease in the AGN luminosities, since z=1, which can be either explained by the declining accretion of the central SMBH in all galaxies \citep[the so-called mass-starvation; i.e.][]{Barger05}, or due to the starvation of the massive galaxies, as the star formation becomes dominated by the lower stellar mass ones \citep[i.e.][]{Cowie96}. Yet, a decrease in the AGN luminosities  (f$_{AGN}$) was also what we observed in the galaxies of our sample for a narrow range in stellar mass (see Table~\ref{tab3}), thus confirming that even though their AGN number densities increase, they seem follow the same evolution as galaxies in other environments.

\section{Summary}
In this paper we have presented the first study of the evolution of the nuclear activity in compact groups of galaxies over the past 3 Gyr (z=0.01-0.23). Our sample is also the largest multi-wavelength compact group sample to-date, consisted of 1,770 isolated compact groups (selected using Hickson's criteria), containing 7,417 galaxies.  We classified the nuclear properties of our galaxies using optical emission line diagrams, as well as mid-IR diagnostic methods. We have also fitted their UV-to-mid-IR spectral energy distributions and estimated the physical properties of the galaxies, which were necessary for our study. Our main conclusions are the following:

\begin{itemize}
\item We observe an increase of $\sim$15\% of the number of AGN-hosts found in late-type galaxies of the dynamically old groups, accompanied by the corresponding reduction of the circumnuclear star formation as we are observing galaxies towards lower redshifts. Examining their Sy2 galaxy fractions a 10-15\% increase is also observed. 

\item Comparing with local isolated field and interacting pair galaxy samples (found at z below 0.05), we do not find differences between the various environments. Yet, this can be also attributed to the small number statistics of the comparison samples.

\item Using both optical, as well as mid-IR colour classifications, we report the absence of powerful Sy1 nuclei (less than 1\% for the galaxies in our sample). These results are also confirmed by our SED modeling. We suggest that gas depletion, due to past star formation activity and tidal stripping, can lead to low accretion rates onto the super-massive black hole, which in turn will result in the lack of the highly-accreting Sy1 nuclei. The  decreasing accretion rates are also confirmed by measuring the AGN H$\alpha$ luminosities of our galaxies, which were reduced by almost an order of magnitude, over the past 3 Gyr.  

\item An increase of about 15\%, has been observed in the AGN fractions of the early-type galaxies in our sample. Examining the observed increase of the Sy2 and LINER fractions in dynamically young groups, we suggest that it may be caused by the on-going morphological transformation of lenticular into elliptical galaxies.

\item We show that galaxies found in dynamically old groups, where they have most probably experienced a larger number of encounters, are more likely to host an AGN into their nucleus, than in dynamically young groups, at any given stellar mass.

\end{itemize}

\section*{Acknowledgments}
T.B. would like to acknowledge support from the DGAPA-UNAM postdoctoral fellowships. D.D. acknowledges support through grant 107313 from PAIIT-UNAM. V.C. would like to acknowledge partial support from the EU FP7 Grant PIRSES-GA-2012-316788. LC also acknowledges financial support from  the {\sc  thales} project 383549  that is jointly  funded by  the European  Union and  the Greek Government in the framework  of the programme ``Education and lifelong learning''. We also appreciate the very useful comments of the referee which helped improve this paper. This research has made use of data products from: Galaxy Evolution Explorer (GALEX), and ultraviolet space telescope operated by Caltech/NASA, Infrared Science Archive (IRSA/Caltech), a UCLA/JPL-Caltech/NASA joint project, and Sloan Digital Sky Survey (SDSS). T.B. would also like to thank, P. Bonfini, G. Maravelias, A. Maragkoudakis and A. Steiakaki for the inspired discussions.

\bsp

\label{lastpage}

\end{document}